\documentclass[twocolumn]{aastex631}
\usepackage{booktabs}
\usepackage{amsmath}
\usepackage{graphicx}
\usepackage{CJK}
\usepackage{subfigure}
\usepackage{threeparttable}

\submitjournal{ApJ}

\newcommand{\beq}{\begin{equation}}
\newcommand{\eeq}{\end{equation}}
\newcommand{\bea}{\begin{eqnarray}}
\newcommand{\eea}{\end{eqnarray}}
 
\begin{document}

\title{Apparently ultra-long period radio signals from self-lensed pulsar-black hole binaries}
\author{Xinxu Xiao (肖欣旭)}\thanks{E-mail: xiaoxx27@mail2.sysu.edu.cn}
\affiliation{School of Physics and Astronomy, Sun Yat-Sen University, Zhuhai, 519082, P. R. China}
\affiliation{CSST Science Center for the Guangdong-Hongkong-Macau Greater Bay Area, Sun Yat-Sen University, Zhuhai, 519082, P. R. China}
\author{Rong-Feng Shen (申荣锋)}\thanks{shenrf3@mail.sysu.edu.cn}
\affiliation{School of Physics and Astronomy, Sun Yat-Sen University, Zhuhai, 519082, P. R. China}
\affiliation{CSST Science Center for the Guangdong-Hongkong-Macau Greater Bay Area, Sun Yat-Sen University, Zhuhai, 519082, P. R. China}

\begin{CJK*}{UTF8}{gbsn}
\begin{abstract}
Pulsar-black hole (BH) close binary systems, which have not been found yet, are unique laboratories for testing theories of gravity and understanding the formation channels of gravitational-wave sources. We study the self-gravitational lensing effect in a pulsar-BH system on the pulsar's emission. Because this effect occurs once per orbital period for almost edge-on binaries, we find that it could generate apparently ultra-long period (minutes to hours) radio signals when the intrinsic pulsar signal is too weak to detect. Each of such lensed signals, or 'pulse', is composed of a number of amplified intrinsic pulsar pulses. We estimate that a radio telescope with a sensitivity of $10\,\rm mJy$ could detect $\sim$ a few systems that emit such signals in our galaxy. The model is applied to three recently found puzzling long-period radio sources: GLEAM-X J1627, PSR J0901-4046, and GPM J1839-10. To explain their observed signal durations and periods, the masses of their lensing components are estimated as $\sim10^4\,\rm  M_{\odot}$, $\sim4\,\rm  M_{\odot}$ and $10^{3-6}\,\rm  M_{\odot}$, respectively, with their binary coalescence times ranging from a few tens to thousands of years. However, the implied merger rates (as high as $\sim 10^{3-4}\,\rm Myr^{-1}$ per galaxy) and the large period decay rates ($>10^{-8}\,\rm s\,s^{-1}$) tend to disfavour this self-lensing scenario for these three sources. Despite this, our work still provides observational characteristics for self-lensed pulsar-BH binaries, which could help the detection of related sources in the future. Finally, for a binary containing a millisecond pulsar and a stellar-mass BH, the Shapiro delay effect would cause a $\geq10\%$ variation of the profile width for the sub-pulses in such lensed signals.
\end{abstract}


\section{Introduction}
Radio pulsars, after their discovery in 1967 \citep{hewish68}, were soon identified as rapidly rotating and highly magnetized neutron stars (NS). Their observed periods of signals are short and stable, making them one of the most precise clocks in the universe. As a result, pulsar binaries become unique laboratories to study stellar evolution \citep{ferdman20, stairs04}, NS magnetospheric physics \citep{breton12, lomia14} and the strong gravitation near compact objects \citep{kramer06, ransom14}. 
Up to now, more than hundreds of pulsar binary systems have been found. However, among them, few binaries harbour another compact object besides pulsar. For instance, only fifteen double NS systems were detected, and thirteen of them were found to contain millisecond pulsars \citep{pol19}. Moreover, the pulsar-black hole (BH) binaries are even considered the holy grail of pulsar astronomy, but their only observed evidence so far is the gravitational-wave (GW) signals from those NS-BH merge events detected by the LIGO–Virgo detector network \citep{2021ApJ...915L...5A, collaboration2024observation}. 

Here we study a possible scenario, the self-lensing event in a pulsar-BH binary, which may enable the observation of their unique electromagnetic emissions, long before their coalescence. Gravitational lensing is an effect that leads to the deflection and amplification of light rays when they travel through a massive and compact astronomical object. After a century since the first successful observation of the bending of light by Sun \citep{1920RSPTA.220..291D}, the gravitational lensing effect has become an important astrophysical tool in various fields, such as the detection of extra-solar planets \citep{mao91, alcock00, kuang22} and the interpretation of the emission from compact binaries \citep{2020MNRAS.495.4061H, 2021MNRAS.503.1703I}. 

Gravitational lensing is classified into three types: Strong lensing, where multiple images of the background source or significant distortion of the source image are visible; Weak lensing, where the source image is slightly distorted; Micro-lensing, where the source image can not be resolved, but the alignment of the moving lens causes a tentative enhancement of the observed source flux. The self-lensing effect we study here belongs to the type of micro-lensing, where the flux amplification can reach orders of magnitude when the lens and the source are extremely aligned to the line of sight. 

The self-lensing effect was first postulated by \citet{1969ApJ...156.1013T} and has been well developed since then. \citet{meader73} studied the orbital-phase-dependent evolution of the amplification for a few cases of different binary components. A similar work was done by \citet{witt94}, who proposed formulae describing the lensing of an extended source. \citet{rahvar11} built a model describing the flux from lensed main-sequence stars. \citet{gould95} considered the self-lensing in binaries of double white dwarfs (WDs), double pulsars, and double supermassive BHs, respectively. His estimation of their detection rates suggested that the self-lensing signals are unlikely to be detected except for binary pulsar systems. Based on this, \citet{marsh01} focused on the compact binaries consisting of a WD and discussed the detection possibility of self-lensing signals for situations where the sources are outside the Einstein ring. As for pulsar-BH systems, the strong-field effect on the observed pulse arrival times was studied both in the case of Schwarzschild BH \citep{wang09b,wang09} and the Kerr BH case \citep{oscoz97,nampa13}. 

So far the observed self-lensing systems are rare, and only five binaries are confirmed, all of which consist of a WD and a star \citep{kruse14, kawahara18, masuda19}. \citet{kruse14} reported the first detected self-lensing binary KOI-3278. However, KOI-3278 shows a very mild amplification, only enhancing the flux by $0.1\%$, which suggests that the source lies outside the Einstein radius, but even so, the signals can still be detected. There are also other systems with possible self-lensing signals, such as the potential supermassive BH binary reported by \citet{2020MNRAS.495.4061H} and the ones proposed by \citet{2021MNRAS.503.1703I} to explain the observed X-ray quasi-periodic eruption sources. 

This work focuses on the self-lensing effect in pulsar-BH binaries. Considering the time-dependent amplification and the pulsar's emission profile, a simple toy model is used to predict the observational appearance of the lensed signal (\S\ref{sec: toy model}). We find that the radio pulses in such systems would appear with some regularity (\S\ref{sec: Ultra-long period sources}). As an extreme case, for example, the pulsar in a binary that is too far from us, or its emission is intrinsically very weak, will have no pulse to be detected most of the time. However, these pulses would be brightened significantly when the pulsar moves behind its BH companion in an edge-on case. Such a situation would exhibit repeated signals, separated by a dark period that is equal to the binary orbital period, which might appear as an extremely slow pulsar. 

This picture is partly motivated by three recently found puzzling radio sources: GLEAM-X J1627 \citep{hurley22}, PSR J0901-4046 \citep{caleb22} and GPM J1839-10 \citep{hurley23}. 
Among them, PSR J0901-4046 \citep{caleb22} shows variable pulse profiles and has the shortest period $\sim1.3\,\rm min$, which is much longer than a normal pulsar. As for GLEAM-X J1627 \citep{hurley22}, its $\sim18.2\,\text{min}$ period and multiple-peak pulse structure are similar to the observational appearance of a self-lensed pulsar-BH binary that our model predicts. GPM J1839-10 \citep{hurley23} is also characterized by an ultra-long period of $\sim22\,\text{min}$. Moreover, it has been found to have been actively existing for three decades, which makes this source even more mysterious.

This paper is organized as follows. In \S\ref{sec: toy model}, we describe the toy model for the self-lensing and the assumed intensity profile of the pulsar's intrinsic emission. We estimate the number of observable self-lensed pulsar-BH binaries in \S\ref{sec: number}. In \S\ref{sec: Ultra-long period sources}, we predict the appearance of ultra-long period radio signals in an edge-on pulsar-BH binary, apply the model to those recently found sources and check if they favour the self-lensing scenario as their physical origin. Additionally, we consider the Shapiro delay effect and estimate its resulting distortion of the lensed pulse profiles in \S\ref{sec: shapiro delay}. Finally, we summarize the results in \S\ref{sec: conclusion}.

\section{Toy Model and Periodic Flares} \label{sec: toy model}

Our model consists of two parts: the self-lensing effect in a pulsar-BH binary (\S\ref{sec: selflensing}) and the pulsar's intrinsic emission(\S\ref{sec: Pulsar emission}). In the former, we consider the pulsar as a point-like source and present the time-dependent evolution of the amplification. In the latter, we calculate the light curve consisting of self-lensed pulses, assuming the intrinsic pulse profiles are of a Gaussian.

\subsection{Self-lensing in binaries} \label{sec: selflensing}
The fundamental parameters and equations of gravitational lensing have been thoroughly defined and established in previous literature  \citep{schneider92, witt94, pacz96}. The readers can refer to these works for detailed derivation and further insights into the topic.

\begin{figure}
\centering
\includegraphics[width=0.45\textwidth, angle=0]{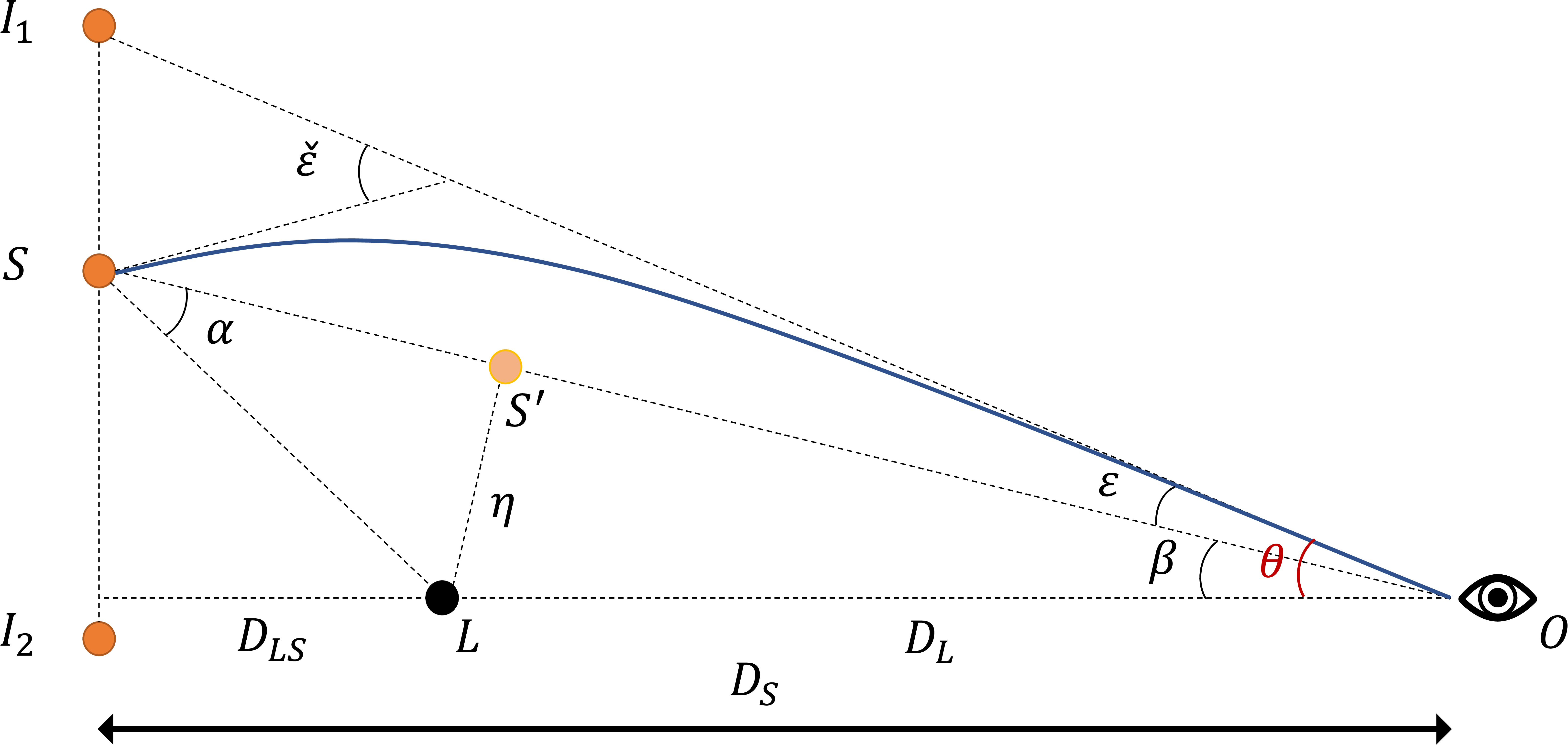}
\caption{The basic geometry in a self-lensing event. The $S$, $L$, and $S'$ denote the source, lens, and the projection of the source in the lens plane, respectively. $I_{\rm 1}$ and $I_{\rm 2}$ are the two images of $S$ distorted by the lens. The distances from the observer to the source and lens are $D_{\rm S}$ and $D_{\rm L}$ respectively. In the case of a binary system, the source and lens are not far apart, $D_{\rm L} \approx D_{\rm S} = D$. The angular separation between the source and lens in the observer's view is represented by $\beta$. As Eq. (\ref{eq: beta}) and (\ref{eq: amppoint}) show, a small value of $\beta$ indicates a significant amplification of lensing.} \label{fig: basicgeo}
\end{figure}
To describe the self-lensing action, a binary system composed of a pulsar and a black hole (BH) with masses in the range of $10\, \rm M_{\odot} < M_{\rm BH} < 10^6\,\rm  M_{\odot}$ is considered. For simplicity, its orbit is assumed to be circular. The binary system has a total mass $M\approx M_{\rm BH}$, an orbital period $P_{\rm b}$, and an orbital separation $a$. Kepler's third law relates these parameters as:
\beq \label{eq: separation}
a = 1.1 \left(\frac{P_{\rm b}}{\rm hr}\right)^{2/3} \left(\frac{M}{10 \rm \,M_\odot}\right)^{1/3}\,\rm R_{\odot}.
\eeq

Below, We will see that strong lensing is possible only when the two components are almost aligned to the line of sight. The Einstein radius of the lens is given by $R_{\rm E} = \sqrt{2 R_{\rm S} a}$, where $R_{\rm S}$ represents the BH Schwarzschild radius. The corresponding angular Einstein radius is $\theta_{\rm E}= \sqrt{2 R_{\rm S} a}/D$, where $D$ denotes the distance to the observer. Assuming that the pulsar is a point-like source, the amplification can be calculated by: 
\beq \label{eq: pointamp}
A(u)= \frac{u^2+2}{u \sqrt{u^2+4}},
\eeq
where 
\beq\label{eq: beta}
u \equiv \frac{\beta}{\theta_{\rm E}},
\eeq 
and $\beta$ is the angular distance between the source and lens in the observer's view. Fig. \ref{fig: basicgeo} illustrates the basic geometry when self-lensing happens.

Eq. (\ref{eq: pointamp}) can be approximated in two asymptotic limits of $u$: 
\beq \label{eq: amppoint}
A(u) \approx \begin{cases}
 &u^{-1},  \mbox{for} ~~ u \lesssim 1, \\
&1+ 2 u^{-4},   \mbox{for} ~~ u \gtrsim 2. 
\end{cases}
\eeq

\begin{figure}
\centering
\includegraphics[width=0.45\textwidth, angle=0]{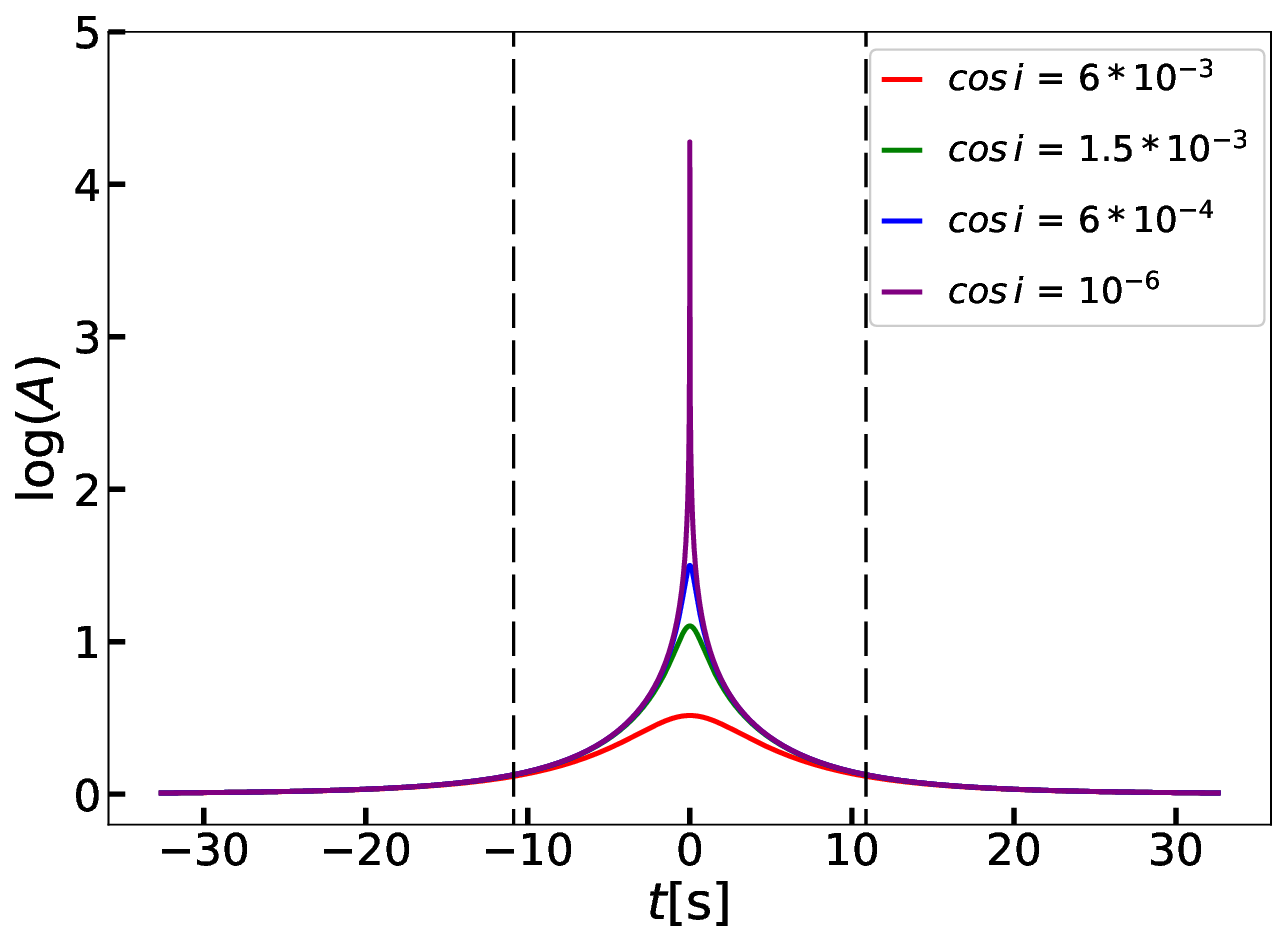}
\caption{The evolution of amplification factor over time in a point-lens and point-source system. The total mass of system is $100\,\rm  M_{\odot} + 1.4\,\rm  M_{\odot}$. The orbital period is $P_{\rm b} = 1\,\rm  hr$. The coloured lines represent different values of $\cos i$, which correspond to various orbital inclinations. $t=0$ corresponds to the moment when the closest lens-source separation is reached. The two vertical dashed lines denote the self-lensing timescale, $t_{\rm lens}$. } \label{fig: Amppoint}
\end{figure}

\begin{figure*}
\centering
\includegraphics[width=0.85\textwidth, angle=0]{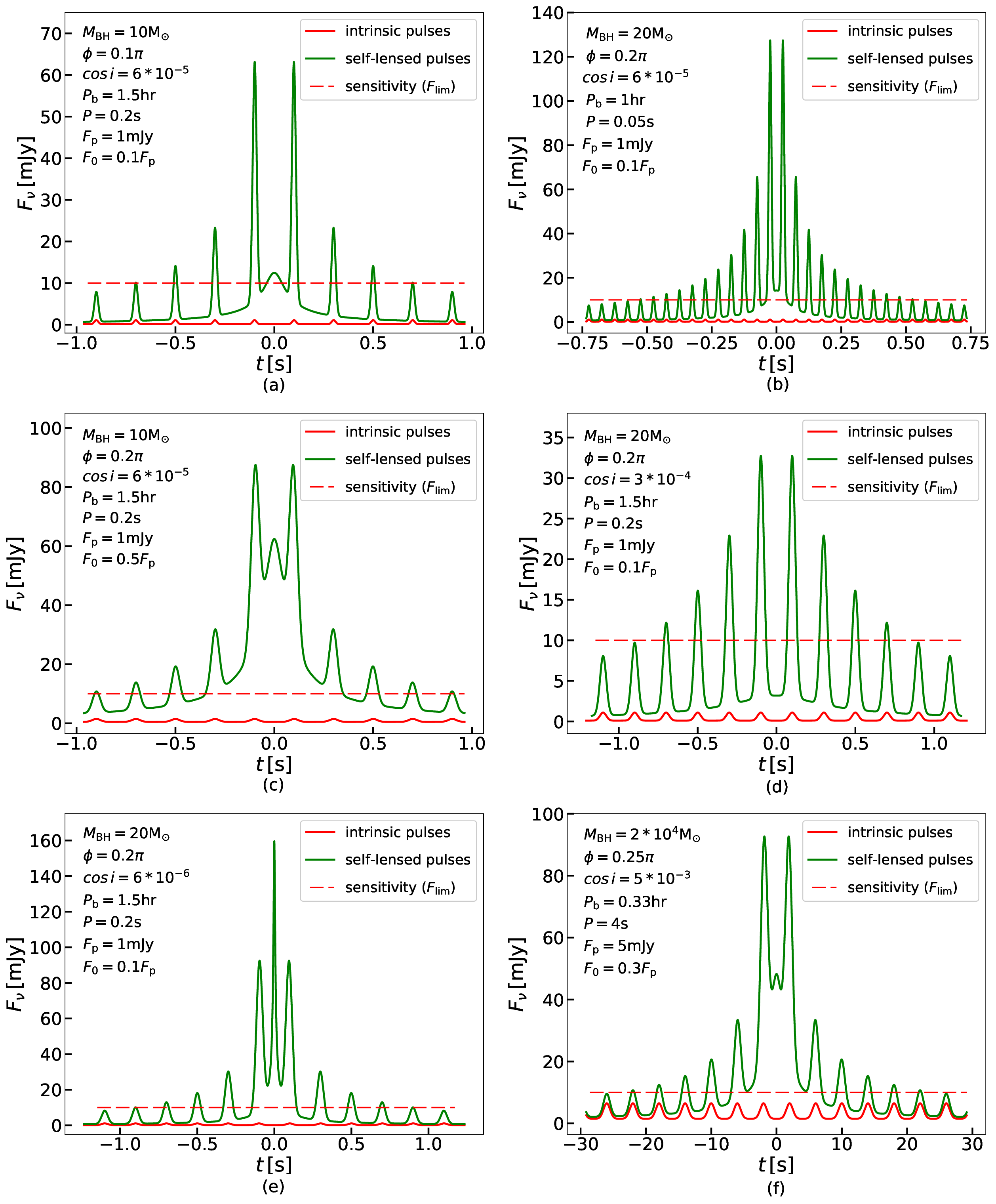}
\caption{The calculated light curves (green solid lines) of the self-lensed pulses with different binary parameters. Except that the NS mass is fixed as $1.4\rm M_{\odot}$, the values of parameters are shown in the up-left corner of each panel. The light curves of intrinsic emission from the pulsar are also represented by the red solid lines. The red dashed lines correspond to a certain detection sensitivity $F_{\rm lim}$, which is set to $10\rm mJy$. In observation, only when the peak flux of lensed pulses exceeds $F_{\rm lim}$ can they be detected. In these particular examples, we deliberately choose the peak time of the intrinsic pulse to showcase a situation where the maximum amplification occurs between two separate pulses. Therefore, the observed central peaks in (a), (c), (e) and (f) are not from the self-lensed pulses but the significantly amplified quiescent flux.}
\label{fig: fluxpoint}
\end{figure*}

The first limit of $u \lesssim 1$ is used when the source lies within the Einstein ring of the lens, in which the lensing amplification increases with the decreasing $u$. For a binary system, the lensing effect should be the most prominent when the source-lens angular separation is at its minimum: $\beta_{\rm min} = a \cos i /D$, where $i$ represents the inclination angle. This yields: 
\beq \label{eq: umin}
\begin{aligned}
u_{\rm min} &= \frac{\beta_{\rm min}}{\theta_{\rm E}} = \sqrt{\frac{a}{2R_{\rm S}}} \cos i\\
&= 0.11\, \left( \frac{P_{\rm b}}{\rm hr} \right)^{1/3} \left( \frac{M}{10\,\rm M_\odot} \right)^{-1/3} \left(\frac{\cos i}{0.001}\right).
\end{aligned}
\eeq
As Eq. (\ref{eq: amppoint}) demonstrates, to achieve a high amplification, $u_{\rm min}$ must be $\ll 1$. The corresponding amplification is $A_{\rm max} \approx u_{\rm min}^{-1}$. It is evident from Eq. (\ref{eq: umin}) that achieving this requires $\cos i \approx 0$, indicating an extremely edge-on case.

When lensing occurs, the transverse relative speed between the source and lens is $v_{\rm t} = 2\pi a/P_{\rm b}$. One can express $u$ as a function of time: 
\beq \label{eq: ut}
u(t)= \sqrt{u_{\rm min}^2 + \left( 2\frac{t}{t_{\rm lens}}\right)^2 }, 
\eeq
where $t=0$ corresponds to the time when $u_{\rm min}$ is achieved, and the time scale
\beq \label{eq: tlens}
\begin{aligned}
t_{\rm lens}  =  \frac{2R_{\rm E}}{v_{\rm t}}  
  =  10.2 \left(\frac{M}{10\,\rm M_\odot}\right)^{1/3} \left(\frac{P_{b}}{\rm hr}\right)^{2/3} \, \mbox{s}
\end{aligned}
\eeq  
characterizes the duration of self-lensing, which is twice the so-called Einstein ring crossing timescale. Note that, when the system is sufficiently nearby, the lensing effects beyond $t_{\rm lens}$ can still be detected, as discussed in \citet{marsh01} and supported by the discovery of KOI-3278 \citep{kruse14}. 

One can characterize the amplified signals by $A_{\rm max}$ and $t_{\rm lens}$ and derive various examples of the time-dependent evolution of the amplification factor with Eq. (\ref{eq: amppoint}) and (\ref{eq: ut}). Fig. \ref{fig: Amppoint} shows such examples for a binary system consisting of a $100\, \rm M_{\odot}$ BH and a $1.4\, \rm M_{\odot}$ NS for different orbital inclinations.

\subsection{Pulsar emission and self-lensing signals} \label{sec: Pulsar emission}

Radio pulsars radiate continuously as they consume their rotational energy. Due to their rapid spin, observers detect periodic short-duration pulses with the spin period $P$. During a self-lensing event, it becomes possible to observe multiple short pulses with their flux amplified. Considering the compact nature of pulsars, it is reasonable to treat them as point-like sources in self-lensing events. Therefore, the self-lensed flux can be obtained by multiplying the amplification factor with the intrinsic flux of the pulsar.

We use a Gaussian-like curve superposed on a constant quiescent flux level to describe the intrinsic pulse profile:  
\beq \label{eq: Fgauss}
\begin{aligned}
 F&(t) = F_{\rm p}e^{-2[\pi (t-t_{\rm p})/\phi]^2}+F_{\rm 0}\\
 &-\frac{P}{2}\,<\,t-t_{\rm p}\,<\,\frac{P}{2}, 
\end{aligned}
\eeq 
where $\phi$ measures the opening angle of the pulsar emission cone, and $t_{\rm p}$ is the time when the peak flux of the pulse is reached. When the magnetic axis is not aligned with the line of sight, there still exist some magnetic field lines oriented towards the observer. Thus, weak radiation from the pulsar still exists, which would be amplified by the lensing effect. A constant flux $F_{\rm 0}$ is used to account for this. Then the amplified flux is obtained as a function of time: 
\beq \label{eq: Flens}
\begin{aligned}
F_{\rm lens}(t) = A(u) \cdot F(t).
\end{aligned}
\eeq

We plot example light curves for self-lensed pulses in Fig. \ref{fig: fluxpoint} for different values of parameters. The horizontal red dashed line represents a telescope's sensitivity $F_{\rm lim}$. If the unlensed peak flux $F_{\rm p}$ is below this sensitivity in observation, then only the amplified pulses are detectable, which would appear as a signal with a much longer period that is equal to $P_{\rm b}$, resulting in an appearance similar to an ultra-long period pulsar.

With different parameter values, the lensed signals show diverse appearances. In Fig. \ref{fig: fluxpoint}, panels (a) and (c) exhibit the cases in which $M_{\rm BH}=10\rm M_{\odot}$.  In comparison to (a), the sub-pulses in (c) have a higher level of $F_{0}$ and $\phi$, which makes the amplified $F_{0}$ comparable to the peak fluxes of the surrounding lensed sub-pulses and results in a larger duty cycle. For the cases in panels (b), (d) and (e), the BH mass is $20\rm M_{\odot}$. Among them, panel (b) showcases a situation where the pulsar spins faster than in (d) and (e), which enables more sub-pulses to be detected in a self-lensing signal. Therefore, the overall shape of the signal resembles the amplification factor evolution shown in Fig. \ref{fig: Amppoint}. The orbital inclination is adjusted to higher ($cos\, i =3\times10^{-4}$) and lower ($cos\, i =6\times10^{-6}$) levels in panels (d) and (e), corresponding to flatter or steeper signal profiles, respectively. In addition, we present an extreme case in panel (f), where the BH mass $M_{\rm BH} = 2\times10^4\rm M_{\odot}$, the spin period $P=4\rm s$ and the orbital period $P_{\rm b}=0.33\rm hr$.

\section{The number estimation for detection} \label{sec: number}
Here we estimate the number of observable self-lensed pulsar-BH systems within the Milky Way. We first compute the number of galactic NS-BH binaries that can reach a threshold lensing amplification $A_{\rm th}$ in \S\ref{sec: numNB}. This calculation also incorporates constraints from recent NS-BH merger events that LIGO-Virgo detects. Then in \S\ref{sec: numPL}, we estimate the number of galactic self-lensed pulsar-BH binaries assuming a uniform radio luminosity for all pulsars.

\subsection{The binary number estimation for a threshold amplification}
\label{sec: numNB}
We start by estimating the lensing probability $\tau(A_{\rm th})$ for a given threshold amplification $A_{\rm th}$. We multiply it with the binary number in the Milky Way, which is a function of the binary mass $M$ and the orbital period $P_{\rm b}$, and then integrate it over the ranges of feasible $M$ and $P_{\rm b}$. Note that, since normally $M_{\rm BH} \gg M_{\rm NS}$, we will simplify our calculation by using $M_{\rm BH}$ to represent the binary mass $M$ as follows. 

In a pulsar-BH binary system, the probability of a self-lensing event is given by \citep{pacz86,gould95}:  
\beq \label{eq: tau}
\begin{aligned} 
\tau=\frac{\pi R_{\rm E}^2}{4\pi a^2}\, = \,\frac{GM_{\rm BH}}{ac^2}.
\end{aligned}
\eeq
It is also the possibility for $u_{\rm min}\leq1$, which corresponds to that the closest angular separation between the pulsar and BH is smaller than the angular Einstein radius. In observation, a threshold amplification $A_{\rm th}$ is needed to make the lensed pulse flux exceed the telescope's sensitivity and become detectable (i.e., as long as $A_{\rm max}>A_{\rm th}$, signals composed of self-lensed pulses can be observed). This implies: $u_{\rm min} \leq u_{\rm th} = A_{\rm th}^{-1}$. The possibility that this constraint holds is:
\beq \label{eq: taunew}
\begin{aligned} 
	\tau(A_{\rm th}) &= \,\frac{\pi (u_{\rm th}R_{\rm E})^2}{4\pi a^2}\\
         &\approx 2\times10^{-9}\left(\frac{M_{\rm BH}}{10\,\rm M_{\odot}}\right)^{\frac{2}{3}}\left(\frac{P_{\rm b}}{\rm hr}\right)^{-\frac{2}{3}}\left(\frac{A_{\rm th}}{100}\right)^{-2}
\end{aligned}
\eeq

\begin{figure*}
\centering
\includegraphics[width=0.85\textwidth, angle=0]{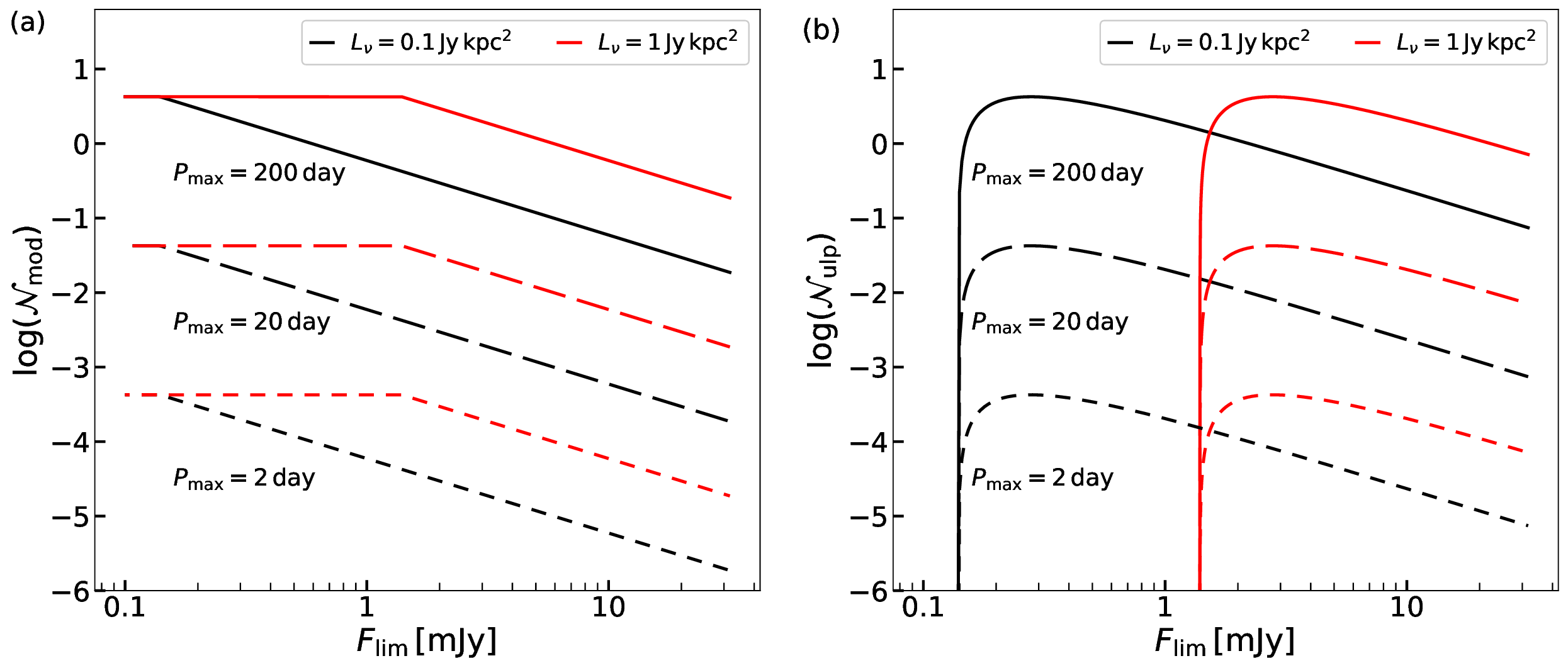}
\caption{The estimated number of observable binaries as a function of sensitivity $F_{\rm lim}$, for given maximum orbital period $P_{\rm max}$ and the pulsar radio pseudo-luminosity $L_{\nu}$, calculated by Eq. (\ref{eq: numpl_d}) and (\ref{eq: numpl}). Other parameters are $M_{\rm BH,min}=10\,\rm M_{\odot}$, $M_{\rm BH,max}=100\,\rm M_{\odot}$ and $P_{\rm min}=30s$. The left and right panels correspond to cases of modulated pulses ($\mathcal{N}_{\rm mod}$) and ultra-long period radio signals ($\mathcal{N}_{\rm ulp}$), respectively. Dashed lines with different dash widths represent different values of $P_{\rm max}$. The pseudo-luminosity $L_{\rm \nu}$ is chosen as $0.1$ and $1\,\rm Jy\,kpc^2$, represented by black and red lines, respectively.}
\label{fig: numpl}
\end{figure*}

The number of binaries in the orbital period interval $dP_{\rm b}$ and the total mass interval $dM$ ($\sim dM_{\rm BH}$) can be written as $dN=\rho(P_{\rm b}, M_{\rm BH})dP_{\rm b}dM_{\rm BH}$. Assuming that the mass distribution of the binary numbers follows the Salpeter mass function:
\beq 
\begin{aligned} 
	p(M_{\rm BH})&\equiv\frac{1}{N_{\rm tot}}\frac{dN}{dM_{\rm BH}}\\
 &=\frac{1-\alpha}{M_{\rm BH,max}^{1-\alpha}\,-\,M_{\rm BH,min}^{1-\alpha}}M_{\rm BH}^{-\alpha},
\end{aligned}
\eeq
and considering that the binaries are only driven to the coalescence stage by the GW emission, the binary population density $\rho(P_{\rm b}, M_{\rm BH})$ can be approximated using the solution of the homogeneous advection equation \citep{christian17, daniel20}: 
\beq \label{eq: rhodpdm}
\begin{aligned} 
	\rho(P_{\rm b}, M_{\rm BH})=\frac{2\mathcal{R}_{\rm i}}{3}\frac{t_{\rm coal}}{P_{\rm b}}p(M_{\rm BH}).
\end{aligned}
\eeq
where $t_{\rm coal}(P_{\rm b}, M_{\rm BH})$ is the coalescence time given by Eq. (\ref{eq: tcoal}), and $\mathcal{R}_{\rm i}$ is the merger rate, which varies depending on the binary type. Because the binary studied here is the NS-BH binary, $\mathcal{R}_{\rm i} = 55\,\rm Gpc^{-3}\,yr^{-1}$ is adopted, based on the recent NS-BH coalescence events detected by LIGO-Virgo \citep{collaboration2024observation}. Assuming that there are $10^{-2}$ Milky-Way-like galaxies per $\rm Mpc^3$ \citep{mont09}, this number could be converted to $\mathcal{R}_{\rm i} \approx 5.5\,\rm MWEG^{-1}\, Myr^{-1}$, where the unit stand for the number of such mergers per Milky Way equivalent galaxy per Myr.

The number of self-lensing binaries with the amplification exceeds $A_{\rm th}$ is estimated as:
\beq \label{eq: NumperMpc}
\begin{aligned} 
	\mathcal{N}_{\rm lens}(A_{\rm th})= 
 \iint\rho(P_{\rm b}, M_{\rm BH})\tau(A_{\rm th})\,dM_{\rm BH}dP_{\rm b}.
\end{aligned}
\eeq
 We set $M_{\rm NS}=1.4\,\rm M_{\odot}$ and allow $M_{\rm BH}$ to range from $M_{\rm BH, min}=10\,\rm M_{\odot}$ to $M_{\rm BH, max}=100\,\rm M_{\odot}$. To identify such systems, a sufficient number of signals must be observed. Therefore, we choose $P_{\rm min}=30\,\rm s$. This ensures that the binaries considered could all exist at least $\sim 1\rm yr$ until coalescence. Because the fraction of binaries with a short period ($\sim \rm min$) is small, $P_{\rm max}$ is set to be $\gg P_{\rm min}$. Thus Eq. (\ref{eq: NumperMpc}) can be written as:
\beq \label{eq: NumGalatic}
\begin{aligned} 
	\mathcal{N}_{\rm lens}(A_{\rm th})\approx 12\times \left(\frac{P_{\rm max}}{200\,\rm day}\right)^2\left(\frac{A_{\rm th}}{10}\right)^{-2}.
\end{aligned}
\eeq
This indicates that within the Milky Way, approximately 12 NS-BH binaries could satisfy the threshold amplification of $A_{\rm th}=10$.

The above outcomes are based solely on the geometric aspect that the binary orbital inclination enables the lensing amplification to reach $A_{\rm th}$. However, considering observational effects (such as signals remaining undetectable even after being amplified) may yield a significant reduction in the observable number, compared with the deduced. We will explore this in the next subsection. 

\subsection{The number of observable self-lensed pulsar-BH binaries}
\label{sec: numPL}

Assuming a uniform opening angle $\phi$ of the pulsar emission cone that is randomly oriented, the probability that a neutron star is observed as a pulsar can be approximated as:
\beq \label{eq: fp}
\begin{aligned} 
	f_{\rm c} =\frac{\Omega_{\rm c}}{4\pi}= \frac{1}{4}\,\sin^2(\phi).
\end{aligned}
\eeq
$\phi=\pi/12$ is set hereafter.

For a particular chromatic pseudo-luminosity, defined as $L_{\nu}=4\pi F_{\nu}D^2$ \citep{lorimer12}, there is a maximum distance beyond which the flux of received signals is smaller than the telescope sensitivity $F_{\rm lim}$, thus being undetectable: $D_{\rm lim} = \left[L_{\nu}/(4\pi F_{\rm lim})\right]^{1/2}$.
To make lensed pulsar emissions observable, the threshold amplification $A_{\rm th}$ must exceed the ratio between the sensitivity and the un-lensed peak flux of pulses: 
\beq\label{eq: ath}
\begin{aligned} 
	A_{\rm th}(D) = \frac{F_{\rm lim}}{F_{\nu}} = \left(\frac{D}{D_{\rm lim}}\right)^2.
\end{aligned}
\eeq
Since the typical radio pseudo-luminosity of a pulsar lies in the range of $(10^{-4}-10)\,\rm  Jy\,kpc^2$ \citep[figure 4]{hurley22}, to ease the calculation, below we simply assume a uniform luminosity for all pulsars, which corresponds to a uniform $D_{\rm lim}$. 

The pulsar emissions from pulsar-BH binaries within $D_{\rm lim}$ can be always detected and exhibit the characteristics same as normal pulsars, except that their flux is modulated by the amplification. Assuming the binaries are isotropically distributed in the Milky-way, their galactic number per orbital period per stellar mass is:
\beq 
\begin{aligned} 
	\rho_{\rm mod}(P_{\rm b}, M_{\rm BH})=f_{\rm c}\left(\frac{D_{\rm lim}}{D_{\rm 25,M}}\right)^2\rho(P_{\rm b}, M_{\rm BH}),
\end{aligned}
\eeq
where $D_{\rm 25, M}=26.8\,\rm kpc$ is the isophotal diameter of the Milky Way and we take it as the size of our galaxy \citep{goodwin98}. 
Thus, the number of binaries $\mathcal{N}_{\rm mod}$ with such modulated pulses can be calculated in a way similar to Eq (\ref{eq: NumperMpc}). Because the intrinsic pulsar emissions can be directly observed, $A_{\rm th}$ becomes a free parameter. We set $F_{\rm lim}=10\,\rm mJy$, $L_{\nu}=1\, Jy\, Kpc^2$ and keep the values of all other parameters the same as in \S\ref{sec: numNB}. The outcome is: 
\beq \label{eq: numpl_d}
\begin{aligned} 
 &\mathcal{N}_{\rm mod}\,= 
 \iint f_{\rm c}\left(\frac{D_{\rm lim}}{D_{\rm 25,M}}\right)^2\rho(P_{\rm b}, M_{\rm BH})\tau(A_{\rm th})\,dM_{\rm BH}dP_{\rm b}\\
 &\approx0.7\left(\frac{P_{\rm max}}{200\,\rm day}\right)^2\left(\frac{A_{\rm th}}{2}\right)^{-2}\left(\frac{F_{\rm lim}}{10\,\rm mJy}\right)^{-1}\left(\frac{L_{\nu}}{1\,\rm Jy\,kpc^2}\right).
\end{aligned}
\eeq

On the other hand, for those pulsars in binaries located beyond $D_{lim}$, their intrinsic emissions are undetectable. Therefore, they become possible candidates for ultra-long period radio sources. Combining Eq.(\ref{eq: taunew}) and (\ref{eq: ath}), the probability $\tau$ becomes a function of distance:
\beq
\begin{aligned} 
	\tau(D) \approx 2\times10^{-5}\left(\frac{M_{\rm BH}}{10\,\rm M_{\odot}}\right)^{\frac{2}{3}}\left(\frac{P_{\rm b}}{\rm hr}\right)^{-\frac{2}{3}}\left(\frac{D}{D_{\rm lim}}\right)^{-4}.
\end{aligned}
\eeq
Using $\rho(P_{\rm b}, M)$ in a distance interval $dD$:
\beq
\begin{aligned} 
	d\rho(P_{\rm b}, M_{\rm BH})=\frac{2f_{\rm c} D}{(D_{\rm 25,M})^2}\rho(P_{\rm b}, M_{\rm BH})dD,
\end{aligned}
\eeq
the number of the ultra-long period radio sources is: 
\beq \label{eq: numpl}
\begin{aligned} 
 &\mathcal{N}_{\rm ulp} =\iiint \frac{2f_{\rm c} D\tau(D)}{(D_{\rm 25,M})^2}\rho(P_{\rm b}, M_{\rm BH})\,dM_{\rm BH}dP_{\rm b}dD\\
 &\approx2.5\times\left(\frac{P_{\rm max}}{200\,\rm day}\right)^2\left(\frac{F_{\rm lim}}{10\,\rm mJy}\right)^{-1}\left(\frac{L_{\nu}}{1\,\rm Jy\,kpc^2}\right).
\end{aligned}
\eeq

It should be noted that our estimation is only done in the Milky Way. Observing such signals from other galaxies requires high amplification due to the significant distances involved, which corresponds to an extremely edge-on case, making the estimated number significantly decrease. For instance, setting $L_{\nu}=1\,\rm Jy\,kpc^2$, a telescope of $F_{\rm lim}=10\,\rm mJy$ could only observe the systems in M31 with $A_{\rm th} \geq 5\times10^3$. Whereas, $\tau(5\times10^3)/\tau(10)\approx4\times10^{-6}$. Thus, the range of $D$ during this integration is limited between $D_{\rm lim}$ and $D_{\rm 25, M}$.

In Figure \ref{fig: numpl}, we explore the correlation between the estimated numbers ($\mathcal{N}_{\rm mod}$ and $\mathcal{N}_{\rm ulp}$) and different values of $P_{\rm max}$, $L_{\nu}$ and $F_{\rm lim}$. Notably, when $F_{\rm lim}$ is at a moderate level ($\gtrsim 0.1\,\rm mJy$ for black lines or $\gtrsim 1\,\rm mJy$ for red lines), these two numbers can be effectively described by the power-law function of the free parameters, as shown in the second lines of Eqs. (\ref{eq: numpl_d}) and (\ref{eq: numpl}). However, when the sensitivity reaches a level at which the telescope can detect the intrinsic pulsar emission from every pulsar-BH binary in the Milky Way, the observed signals will manifest as modulated pulsar signals, and no puzzling ultra-long period sources can be found. Consequently, $\mathcal{N}_{\rm mod}$ becomes steady and $\mathcal{N}_{\rm ulp}$ sharply drops. 

As Eqs. (\ref{eq: numpl_d}) and (\ref{eq: numpl}) show, if all the pulsar's radio luminosity is $1\,\rm  Jy\,kpc^2$, there will be approximately $3$ pulsar-BH binaries in the Milky Way with observable self-lensing signals that can be detected by a telescope with a sensitivity of $10\,\rm mJy$. Among them, about $2$ binaries would exhibit ultra-long period characteristics and the rest are only modulated pulsar signals. The numbers $\mathcal{N}_{\rm mod}$ and $\mathcal{N}_{\rm ulp}$ are sensitive to $P_{\rm max}$, i.e., $\propto P_{\rm max}^2$, as shown in Eqs. (\ref{eq: numpl_d}) and (\ref{eq: numpl}). This suggests that the orbital periods of these binaries should be long, ranging from tens to hundreds of days. 

\begin{figure*}
\centering
\includegraphics[width=0.6\textwidth, angle=0]{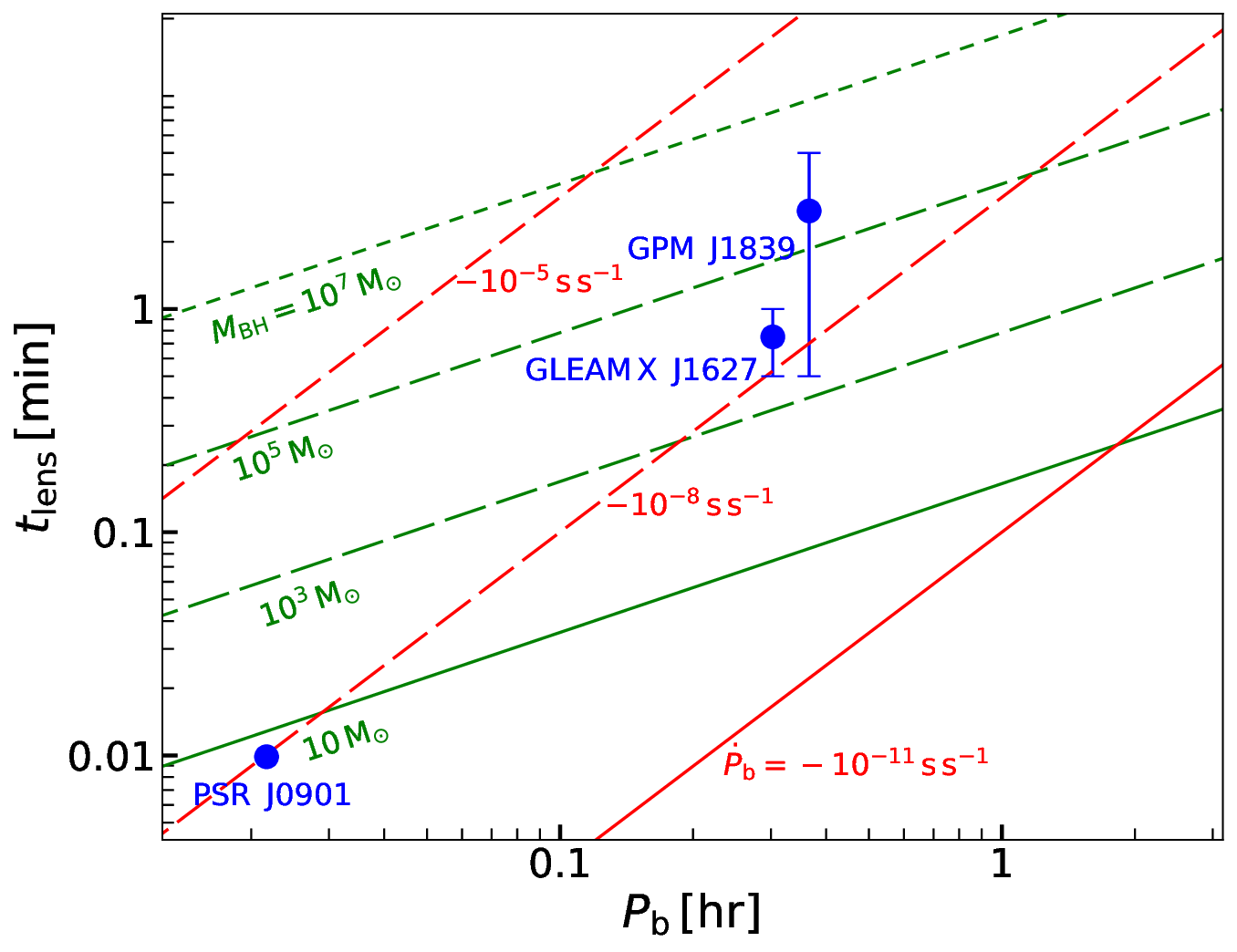}
\caption{The predicted signal durations and periods for the ultra-long period radio sources that are due to the self-lensed pulsar-BH binaries. Here we take $t_{\rm lens}$ given by Eq. (\ref{eq: tlens}) as the duration of the signals and $P_{\rm b}$, the binary orbital period, as the signal period. The green and red lines are for the constant BH mass $M_{\rm BH}$ and decay rate of orbital period $\dot{P_{\rm b}}$, respectively. The recently found ultra-long period radio sources: GLEAM-X J1627, PSR J0901-4046, and GPM J1839-10 are also plotted and marked as blue symbols.}
\label{fig: pspace}
\end{figure*}

\section{Ultra-long period radio sources} \label{sec: Ultra-long period sources}

In real observations, the self-lensed pulsar signals can exhibit two different signatures, depending on whether the pulsar's intrinsic emission is detectable:  

In binaries where the pulsar's intrinsic radio luminosity is sufficiently bright, the observed signal would resemble that of a typical radio pulsar, except that the flux of pulses is modulated by the lensing effect, with the peak flux periodically rising and falling in every orbital period.

On the other hand, when the pulsar's intrinsic emission is too weak to be detected, the observer would only see the signals consisting of the self-lensed pulses that have peak flux above the sensitivity. These pulses are emitted near $t=0$ and appear in every binary orbital period, forming a signal with a multiple-peak structure, as the several peaks that surpass the red dashed line shown in Fig. \ref{fig: fluxpoint}. Such a phenomenon is most likely to occur in binaries that are sufficiently edge-on because significant amplification is needed. Since normally $P_{\rm b}\gg P$, this phenomenon offers a possible explanation for the origin of those enigmatic radio sources with extremely long periods.

Compared to close binaries, such lensed signals at super-long periods from wide binaries might be difficult to identify and track in observations. However, when the binaries are too close, the lifetime of these signals may be short due to the final coalescence driven by GW emission.
In Fig. \ref{fig: pspace}, we explore the parameter space of $t_{\rm lens}$ and $P_{\rm b}$ for the ultra-long period self-lensed signals, where $t_{\rm lens}$ and $P_{\rm b}$ can be identified as the duration of the signals and the ultra-long period of the signals, respectively. The green and red lines represent the apparent pulse 'durations' and periods for the constant BH mass $M_{\rm BH}$ and constant orbital period decay rate $\dot{P_{\rm b}}$, respectively. Here $\dot{P_{\rm b}}$ are determined from the coalescence time $t_{\rm coal}$ as in \citet{peters63}:
\beq \label{eq: pbdot}
\begin{aligned}
\dot{P_{\rm b}} = -\frac{3}{8}\frac{P_{\rm b}}{t_{\rm coal}} .
\end{aligned}
\eeq
and
\beq \label{eq: tcoal}
\begin{aligned}
t_{\rm coal} & = \frac{5}{256}(2\pi)^{-8/3}\frac{c^5}{G^{5/3}}\frac{(M_{\rm BH}+M_{\rm NS})^{1/3}}{M_{\rm BH}M_{\rm NS}}P_{\rm b}^{8/3}\\
&\approx 1.3\times10^{4}\left(\frac{M_{\rm BH}}{10\,\rm M_{\odot}}\right)^{-2/3}\left(\frac{P_{\rm b}}{10\,\rm min}\right)^{8/3} \rm yr,
\end{aligned}
\eeq 
where the last step of Eq. (\ref{eq: tcoal}) used $M_{\rm BH}\gg M_{\rm NS} \equiv 1.4M_{\odot}$. The three ultra-long period radio sources are also plotted in the figure and will be discussed as follows.

\subsection{Puzzling radio sources} \label{sec: puzzling radio sources}

When carrying out this work, we came across three recently discovered interesting radio sources, namely GLEAM-X J1627 \citep{hurley22}, PSR J0901-4046 \citep{caleb22} and GPM J1839-10 \citep{hurley23}. They all exhibit characteristics similar to extremely slow-rotating pulsars with spin periods of approximately $18.2\,\rm min$, $1.3\,\rm min$, and $22.0\,\rm min$, respectively.

Among them, PSR J0901-4046 \citep{caleb22} stands out with the shortest period ($1.3\,\rm min$), with its pulse width about $0.3\,\rm s$, indicating the pulse duration is around $0.6\,\rm s$. Its pulses present a diverse appearance of profiles and are classified into seven distinct morphology types that range from partial nulling to asymmetric quasi-periodic peaks. Notably, this source has been actively radiating since its discovery. With the period derivative tightly constrained to $\dot{P} \approx 2.25\times10^{-13}\,\rm s\,s^{-1}$, its position on the $P-\dot{P}$ diagram can be placed above the so-called death line in several pulsar-emission mechanism models. Additionally, the snapshot images from the MeerTRAP data reveal a diffuse, shell-like structure surrounding the source, reminiscent of the environment in supernova remnants \citep{caleb22}. 

In contrast, GLEAM-X J1627 \citep{hurley22} showcases signals with a much longer period ($18.2\,\rm min$) and a relatively stable pulse morphology, which exhibits a multi-peak structure. Its pulse duration ranges from $30$ to $60\,\rm s$, and the period of $18.18\,\rm min$ poses challenges to existing pulsar models. Even more puzzling, after three months of activity, it disappeared from observations and could not be found in available data. This ultra-long period but short-lived behaviour has prompted a range of interpretations, including a WD with relatively strong magnetic fields \citep{katz22, loeb22} and a spinning-down magnetar \citep{beniamini23,suv23,tong23}.

GPM J1839-10 \citep{hurley23} exhibits similar long-period ($22.0\,\rm min$) characteristics to GLEAM-X J1627. Its pulse duration varies significantly, ranging from $30$ to $300\,\rm s$. What truly sets it apart is its prolonged existence, having been actively observed for three decades. This extended activity defies those magnetar models that have been suggested for the short activity of GLEAM-X J1627 \citep{suv23,tong23}.

An alternative interpretation is an isolated and highly magnetized WD, which could explain the long periods due to its larger moment of inertia than NSs. However, as \citet{rea23} have shown, the locations of GLEAM-X J1627 and GPM J1839-10 in the $P$-$B$ plane (i.e., high $P$ and low $B$) imply relatively weak surface dipolar magnetic fields for them, which makes it still challenging for a magnetized WD to produce signals as bright as those observed. As the conventional model for normal pulsars is already ruled out due to the ultra-long period, exploring alternative physical origins becomes necessary, especially for GLEAM-X J1627 and GPM J1839-10.

Therefore, we apply our model to these sources to check if they are actually self-lensed pulsars. In Fig. \ref{fig: pspace}, PSR J0901-4046 is represented just by a blue dot due to its well-constrained pulse duration. For GLEAM-X J1627 and GPM J1839-10, we plot them as blue dots with error bars, where the dots correspond to the median value of their pulse durations and the bars represent a range between the maximum and minimum values. The results suggest BH masses of $4\, \rm M_{\odot}$ and $\sim10^{4}\, \rm M_{\odot}$ for PSR J0901-4046 and GLEAM-X J1627, with corresponding coalescence times of $10^{2}\,\rm yrs$ and $\sim6\times10^{2}\,\rm yrs$, respectively. 
For GPM J1839-10, its variable signal duration makes the results range in several orders of magnitudes: The BH mass ranges from $\sim10^3$ to $\sim10^6\,\rm M_{\odot}$, with coalescence times spanning from $\sim30$ to $\sim3\times10^{3}\,\rm yrs$. 

In our estimation, the coalescence time derived for GLEAM-X J1627 appears excessively long, inconsistent with its reported three-month active period. Conversely, PSR J0901-4046 has been staying active since its detection, which aligns well with the estimated 100-year lifetime. However, the derived mass for its lens is $\sim4\,\rm  M_{\odot}$, which lies in the $2-5\,\rm M_{ \odot}$ mass gap and is not typical for observed stellar-mass BHs so far. Regarding GPM J1839-10, the derived coalescence time range ($\sim30$ to $\sim3\times10^{3}\,\rm yrs$) aligns with its three-decade activity, implying monitoring a continuing activity of the system for at least another 30 years, which can be verified by the future monitoring.

An observationally more meaningful parameter is the theoretically predicted decay rate of the orbital period $\dot{P_{\rm b}}$, which can be verified by the derivative of the signal period $\dot{P}$. We estimated it for the three sources to be: $\dot{P_{\rm b}} \approx -10^{-8}\,\rm s\,s^{-1}$ and $\approx-2\times10^{-8}\,\rm s\,s^{-1}$, for PSR J0901-4046 and GLEAM-X J1627, respectively. For GPM J1839-10, it is from $-5\times10^{-9}\,\rm s\,s^{-1}$ to $-5\times10^{-7}\,\rm s\,s^{-1}$.

However, our model predicts negative, large period derivatives ($\sim10^{-8}\,\rm s\,s^{-1}$), apparently inconsistent with the reported $\dot{P}$ of the three sources, especially for PSR J0901-4046: It has a positive $\dot{P} \approx 2.25\times10^{-13}\,\rm s\,s^{-1}$ \citep{caleb22}. The other two sources, GLEAM-X J1627 and GPM J1839-10, have reported $\dot{P}$ below two upper limits: $\leq 1.2\times10^{-9}\,\rm s\,s^{-1}$ \citep{hurley22} and $\leq 3.6\times10^{-13}\,\rm s\,s^{-1}$ \citep{hurley23}. 

Assuming the self-lensing scenario we describe above, GLEAM-X J1627 and GPM J183910 are interpreted as intermediate-mass ratio inspirals (IMRIs), consisting of an NS and an intermediate-mass black hole (IMBH). One of the most promising environments to embed IMBHs is the centres of star clusters. Since the formation of an IMBH is related to the numerous low-mass compact objects that orbit around it, star clusters are also ideal places to find IMRIs. 

Many works studied the formation and the evolving fate of IMRIs. \citet{mandel08} discussed four mechanisms that could form IMRIs, including the hardening of a CO-IMBH Binary through three-body interactions, Kozai resonance, direct capture via GW emission and inspiral of a CO from a tidally captured main-sequence star. Among them, the three-body interaction between an IMBH and a binary system is recognized as the primary pathway. Based on this, the estimated formation rate of the NS-IMBH IMRI systems that are detectable in GWs at the coalescence stage would be $\sim 0.1\,\rm Gal^{-1}\, Myr^{-1}$ \citep{mandel08}. This result was supported by N-body simulations \citep{kons13, MacLeod_2016, Fragione_2018}.

\vspace{5pt}
\begin{table}
\centering
\begin{tabular}{c c c c}
 \toprule[1.2pt]
  & $M_{\rm BH}$&$t_{\rm coal}$&$\mathcal{R}_{\rm NSBH}$\\
 & $(\rm M_{\odot})$& $(\rm yrs)$&$\rm(MWEG^{-1}\,Myr^{-1})$\\ \hline 
 PSR J0901& $4$&$~100$&$10^4$\\ \hline  
 GLEAMX J1627& $10^4$&600&$1.7\times10^3$\\ 
 \hline 
 GPM J1839& $10^{3-6}$& $3\times10^{1-3}$&$3.3\times10^{2-4}$\\
 \bottomrule [1.2pt]
\end{tabular}
\caption{The BH masses, coalescence times and merger rates for the three sources, derived by applying the self-lensing pulsar-BH scenario to them.}
\label{table: three sources}
\end{table}

If these three sources were indeed self-lensed pulsar-BH binaries, we could roughly derive the underlying merger rates for each of them based on their estimated $t_{\rm coal}$ (see Table \ref{table: three sources}):
\beq \label{eq: restimated}
\begin{aligned}
\mathcal{R}_{\rm NSBH} \approx 10^4\times \left(\frac{t_{\rm coal}}{100\,\rm yrs}\right)^{-1}\,\rm MWEG^{-1}\,Myr^{-1},
\end{aligned}
\eeq
where $\rm MWEG$ stands for Milky Way equivalent galaxies, as these sources are found within the Milky Way. On the other hand, the LIGO-Virgo detection of NS-BH mergers gives a merger rate estimate of $\mathcal{R}_{\rm i} = 5.5\,\rm MWEG^{-1}\,yr^{-1}$ (see \S\ref{sec: numNB}), which is much lower than $\mathcal{R}_{\rm NSBH}$, especially for PSR J0901-4046.

Note that for GLEAM-X J1627 and GPM J1839-10, their corresponding $\mathcal{R}_{\rm NSBH}$ cannot be directly compared to $\mathcal{R}_{\rm i}$, since the GWs from a NS’s inspiral into an IMBH is currently difficult to detect. However, as mentioned above, the anticipated event rate of GW-observable NS-IMBH IMRI is just $\sim 0.1\,\rm Gal^{-1}\, Myr^{-1}$ \citep{mandel08}. Additionally, \citet{Fragione_2018} estimated that there would be $\sim 10^3$ IMBHs within $1\,\rm kpc$ from the Galactic centre, and the IMBH-SBH merger rate in the local universe is $\sim\,0.1-1\,\rm MWEG^{-1}\, Myr^{-1}$. Though NSs may be more common than SBHs, it is hard to imagine that the corresponding merger rate could be higher by several orders of magnitude.

Therefore, we conclude that these three puzzling radio sources are unlikely the self-lensed pulsars, because the predicted period decay rates exceed the observational constraints, and the implied merger rates are too high.

\section{The Shapiro delay} \label{sec: shapiro delay}
The Shapiro delay, also known as gravitational time delay, is the reduction of the equivalent speed of light under the influence of the gravitational potential along the light's path, which leads to the delay in the arrival time of photons \citep{shapiro64}. Here we investigate whether it will cause a detectable distortion of the self-lensed pulse profiles in a pulsar-BH system. 

In the case of an exact edge-on binary system, utilizing the simplified equation by \citet{possel19}, the phase-dependent Shapiro delay within one orbital period can be expressed as: 
\beq \label{eq: t_shapiro}
\begin{aligned} 
	t_{\rm sd} \approx \frac{2GM}{c^3} \text{ln} \left[ \frac{2D}{a}\frac{1}{1-\cos{(2\pi\Phi_{\rm t})}} \right],
\end{aligned}
\eeq 
where $\Phi_{\rm t}$ denotes the orbital phase. $\Phi_{\rm t}=0$ corresponds to $u=u_{\rm min}$, and $\Phi_{\rm t}=0.5$ corresponds to when the source moves in the exact front of the lens.  

As evident from Eq. (\ref{eq: t_shapiro}), the Shapiro delay is solely dependent on the BH mass and the pulsar's current orbital phase. Even though pulsars are fast-rotating, their position would still undergo a slight shift after completing one spinning due to orbital motion, which corresponds to a change in the orbital phase. Consequently, even photons emitted within the same pulse will encounter different values of Shapiro delays. This difference in the delay is generally marginal compared to the spin period $P$. However, pulses are significantly amplified only when the BH is massive or the source and lens are nearly aligned with the line of sight. In such a case, the difference between the delays of the first and last photons within a single pulse could be on par with the pulsar's spin period $P$. This would lead to the distortion of pulse profiles.

\begin{figure}
\centering
\includegraphics[width=0.45\textwidth, angle=0]{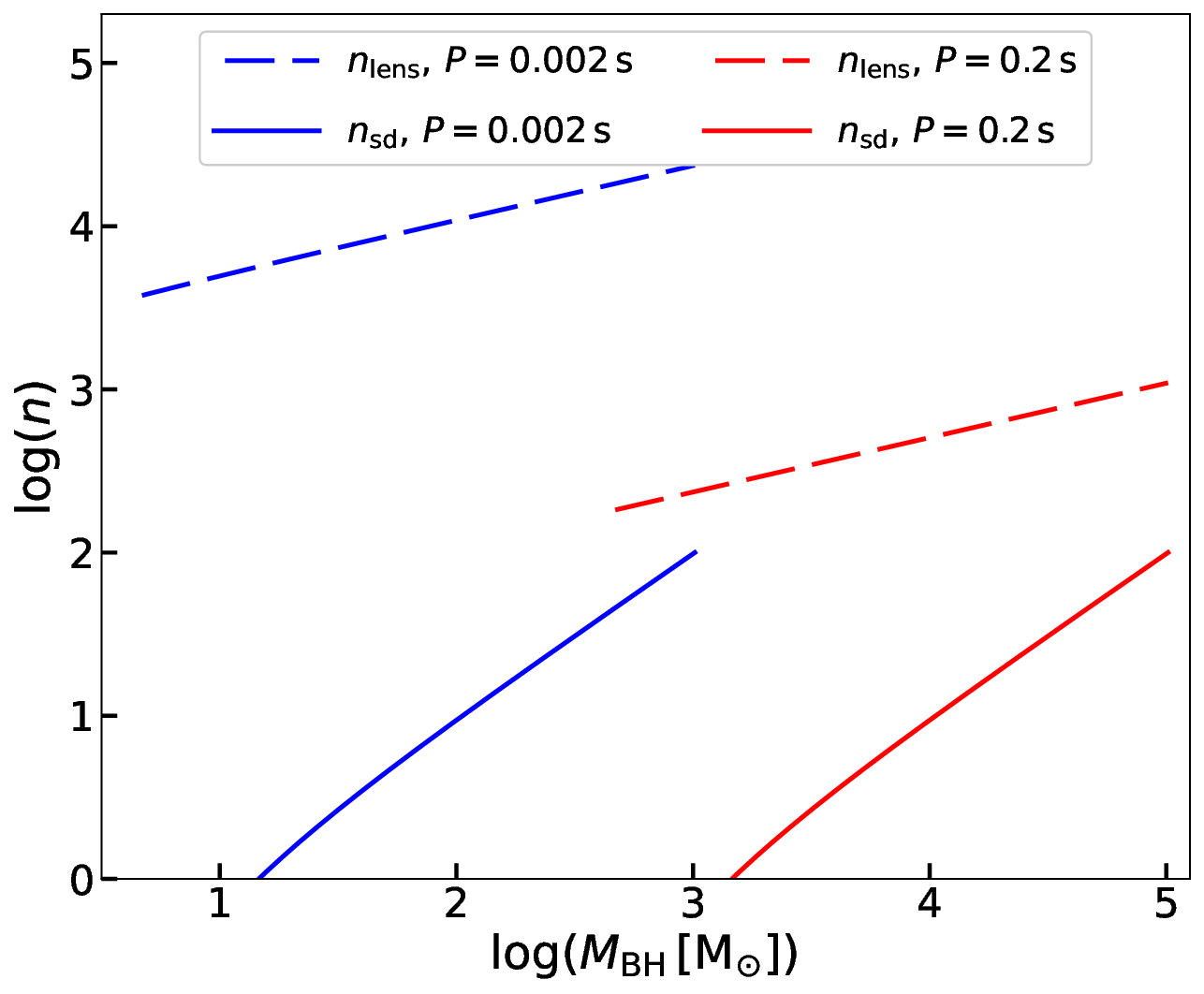}
\caption{The dependence of the number of pulses that are significantly affected by Shapiro delay (for solid lines) and self-lensing effect (for dashed lines) as a function of BH mass. The blue and red lines correspond to the situation when the pulsar is a millisecond pulsar ($P=0.002\,\rm s$) or a normal pulsar ($P=0.2\,\rm s$), respectively.}
\label{fig: delshapiro_p}
\end{figure}

Such difference is measured by: 
\beq
\Delta t_{\rm sd}(\Phi_{\rm t})=t_{\rm sd}(\Phi_{\rm t})-t_{\rm sd}(\Phi_{\rm t}+P/P_{\rm b}).
\eeq
If $\Delta t_{\rm sd}$ is comparable to or larger than the pulse duration or the spin period, it results in significant distortions that manifest as the narrowing or widening of the pulse duration and the time interval between pulses, depending on the orbital phase. If the pulses are emitted at an orbital phase $\Phi_{\rm t} < 0$, the narrowing effect is observed, whereas the widening effect occurs for pulses emitted at $\Phi_{\rm t} > 0$. Furthermore, if a pulse coincides with orbital phase $\Phi_{\rm t} = 0$, the moment of maximum amplification, its peak will be shifted to a later arrival time, deviating from its intrinsic symmetric shape.

Defining a distortion of at least $10\%$ ($\Delta t_{\rm sd}(\Phi_{\rm t})\geq0.1\, P$) as the significant shape alteration, the threshold orbital phase $\Phi_{\rm th}$ can be derived, within which this distortion criterion holds. The number of pulses that the pulsar can emit when moving from $-\Phi_{\rm th}$ to $\Phi_{\rm th}$ is defined as $n_{\rm sd}=2\Phi_{\rm th} P_{\rm b}/P$. We plot it as a function of $M_{\rm BH}$ in Fig. \ref{fig: delshapiro_p}. The blue and red lines correspond to the millisecond and normal pulsar cases. $n_{\rm lens}=t_{\rm lens}/P$, which is the number of pulses that are affected by the self-lensing effect, is also plotted as dashed lines for comparison. A relatively short orbital period $P_{\rm b} = 1\,\rm h$ is chosen, which represents an extreme case compared to most binary systems, and it results in a short $t_{\rm lens}$ and a small $n_{\rm lens}$.

We find that for pulsar-BH systems with a normal pulsar, the effect of Shapiro delay is negligible unless $M_{\rm BH}>10^3\,\rm  M_{\odot}$. However, for systems with a millisecond pulsar, a stellar-mass BH is enough to change the shape of pulse profiles. Because $n_{\rm lens}$ is much larger than $n_{\rm sd}$, even for the $P_{\rm b}=1\,\rm h$ case, the time delay affects only a small fraction of the self-lensed pulses. On the other hand, for systems with massive BHs, $n_{\rm sd}$ is still considerable. As Fig. \ref{fig: delshapiro_p} shows, the value of $n_{\rm sd}$ varies from several to hundreds, making this effect measurable.

\section{Discussion and Conclusion} \label{sec: conclusion}
This paper studies the self-lensing effect in pulsar-BH binaries, aiming to predict their unique observational appearance. Using a toy model, we demonstrate that the self-lensing amplifies the pulsar pulses during each orbital period. The self-lensed pulses would appear as radio sources with ultra-long periods when the intrinsic emission of the pulsar is too dim to be detected. In this scenario, because only the self-lensed pulses can be observed, the signal period is the binary orbital period $P_{\rm b}$, which is typically minutes to hours, even $\sim 1\,\rm  day$. The signal duration is the duration for self-lensing or the Einstein ring crossing timescale.

As the self-lensing effect only happens when the binary is edge-on and when the pulsar's emission cone points toward the observer, the number of non-lensed pulsar-BH binaries should be much more than those self-lensed. However, it usually needs a few years of follow-up observation and precise measurements to determine a non-lensed pulsar-BH binary \citep{liu14}. Thus, 
 even if the existing number of the non-lensed systems is large, only a tiny fraction of them can be detected by current observational technique, and it is reasonable that there is no finding of the pulsar-BH binary so far. On the other hand, the self-lensing effect not only makes the signals brighter but also brings them some special observational appearances (see \S\ref{sec: Ultra-long period sources}), which makes them more easily identified than those non-lensed signals.

We estimate that, in the Milky Way, there would be at least $12$ NS-BH binaries that can reach a threshold amplification of $10$. Observationally, there would be $\sim3$ pulsar-BH binaries with self-lensing signals observable for a telescope with a sensitivity of $10\,\rm mJy$, and $\sim2$ of them would exhibit ultra-long period radio signals. Because of the assumption that all pulsars have a uniform radio luminosity, the actual number of observable sources may be even lower due to the existence of numerous candidates with luminosity lower than the value selected. However, telescopes with lower sensitivities ($\sim \rm \mu Jy$ for long period signals), such as FAST \citep{han21} and SKA \citep{braun19}, hold the potential to identify more sources that contain a low luminosity pulsar, increasing the possibility of detecting such systems.

Three recently found long-period radio sources, GLEAM-X J1627 \citep{hurley22}, PSR J0901-4046 \citep{caleb22} and GPM J1839-10 \citep{hurley23} are puzzling about their physical origins. We apply our model to these sources. Assuming they are self-lensed pulsar-BH binaries, our modelling suggests BH masses of $\sim10^{4}\,\rm  M_{\odot}$ and $\sim4\,\rm  M_{\odot}$ for the former two sources, with coalescence times of $\sim6\times10^{2}\,\rm yrs$ and $\sim10^{2}\,\rm yrs$, respectively, basing on their inferred orbital separations. Given the substantial variation of the signal duration, the BH mass of the third source, GBM J1839-10 is estimated to range between $10^3$ and $10^6\, \rm M_{\odot}$, with a coalescence time spanning $\sim 30$ to $\sim3\times10^{3}\,\rm yrs$, implying at least 30 years more of activity, which could be verified by future observation. 

Our model predicts negative, large values ($\sim10^{-8}$ s s$^{-1}$) of the period derivative for the three sources. However, their observed $\dot{P}$'s are all much smaller ($\lesssim 10^{-9}$ s s$^{-1}$) and even favour positive values (slowing down) for the source PSR J0901-4046. Under the self-lensing scenario, this discrepancy could result from other physical processes. For instance, the spin angular momentum of the pulsar might be transferred to the orbital one, via tidal or some sort of magnetic braking effect, thus reversing the orbital period decay. 

However, if the three sources are self-lensed NS-BH binaries, their corresponding merger rates derived from their coalescence times would be high ($\sim 10^4\,\rm MWEG^{-1}\, Myr^{-1}$), much higher than that given by the up-to-date GW observations ($\sim 5.5\,\rm MWEG^{-1}\, Myr^{-1} $). Therefore, we tend to suggest that the self-lensing scenario for the three sources is disfavoured.

Although the implied high merger rate and the observed period derivative of these three sources tend to disfavour the self-lensing model for them, our work still provides distinguishable observation appearances (\S\ref{sec: toy model}) and the number prediction (\S\ref{sec: number}) for those self-lensed pulsar-BH binaries. It still holds the potential to detect the self-lensed pulsar-BH binaries in the future, especially with more sensitive facilities.

We also investigate the impact of Shapiro delay on the self-lensed pulse profiles. We find that in pulsar-BH binaries with a normal pulsar, the effect of the Shapiro delay is negligible unless the BH mass exceeds $10^3\,\rm  M_{\odot}$. However, for systems with a millisecond pulsar, even a stellar-mass BH can alter the shape of the pulse profiles, which would cause at least $10\%$ variation in profile width of the sub-pulses that is emitted inside a threshold orbital phase $\Phi_{\rm th}$, exhibiting either the compression or stretching of the profile.

\section*{}
This work is supported by National Natural Science Foundation of China (grants 12073091, 12261141691 and 12393814). We thank the anonymous reviewer whose comments have helped improve the quality of the manuscript. We also thank the comments from Tam, Pak-Hin Thomas and Long Ji.

\bibliography{self-lensing}

\begin{thebibliography}{}
\expandafter\ifx\csname natexlab\endcsname\relax\def\natexlab#1{#1}\fi
\providecommand{\url}[1]{\href{#1}{#1}}
\providecommand{\dodoi}[1]{doi:~\href{http://doi.org/#1}{\nolinkurl{#1}}}
\providecommand{\doeprint}[1]{\href{http://ascl.net/#1}{\nolinkurl{http://ascl.net/#1}}}
\providecommand{\doarXiv}[1]{\href{https://arxiv.org/abs/#1}{\nolinkurl{https://arxiv.org/abs/#1}}}

\bibitem[{Abbott {et~al.}(2021)}]{2021ApJ...915L...5A}
Abbott, R., {et~al.} 2021, \apjl, 915, L5, \dodoi{10.3847/2041-8213/ac082e}

\bibitem[{{Alcock} {et~al.}(2000){Alcock}, {Allsman}, {Alves}, {Axelrod}, {Becker}, {Bennett}, {Cook}, {Drake}, {Freeman}, {Geha}, {Griest}, {Lehner}, {Marshall}, {Minniti}, {Nelson}, {Peterson}, {Popowski}, {Pratt}, {Quinn}, {Stubbs}, {Sutherland}, {Tomaney}, {Vandehei}, {Welch}, \& {MACHO Collaboration}}]{alcock00}
{Alcock}, C., {Allsman}, R.~A., {Alves}, D.~R., {et~al.} 2000, \apj, 541, 734, \dodoi{10.1086/309484}

\bibitem[{{Beniamini} {et~al.}(2023){Beniamini}, {Wadiasingh}, {Hare}, {Rajwade}, {Younes}, \& {van der Horst}}]{beniamini23}
{Beniamini}, P., {Wadiasingh}, Z., {Hare}, J., {et~al.} 2023, \mnras, 520, 1872, \dodoi{10.1093/mnras/stad208}

\bibitem[{{Braun} {et~al.}(2019){Braun}, {Bonaldi}, {Bourke}, {Keane}, \& {Wagg}}]{braun19}
{Braun}, R., {Bonaldi}, A., {Bourke}, T., {Keane}, E., \& {Wagg}, J. 2019, arXiv e-prints, arXiv:1912.12699, \dodoi{10.48550/arXiv.1912.12699}

\bibitem[{{Breton} {et~al.}(2012){Breton}, {Kaspi}, {McLaughlin}, {Lyutikov}, {Kramer}, {Stairs}, {Ransom}, {Ferdman}, {Camilo}, \& {Possenti}}]{breton12}
{Breton}, R.~P., {Kaspi}, V.~M., {McLaughlin}, M.~A., {et~al.} 2012, \apj, 747, 89, \dodoi{10.1088/0004-637X/747/2/89}

\bibitem[{{Caleb} {et~al.}(2022){Caleb}, {Heywood}, {Rajwade}, {Malenta}, {Stappers}, {Barr}, {Chen}, {Morello}, {Sanidas}, {van den Eijnden}, {Kramer}, {Buckley}, {Brink}, {Motta}, {Woudt}, {Weltevrede}, {Jankowski}, {Surnis}, {Buchner}, {Bezuidenhout}, {Driessen}, \& {Fender}}]{caleb22}
{Caleb}, M., {Heywood}, I., {Rajwade}, K., {et~al.} 2022, Nature Astronomy, 6, 828, \dodoi{10.1038/s41550-022-01688-x}

\bibitem[{{Christian} \& {Loeb}(2017)}]{christian17}
{Christian}, P., \& {Loeb}, A. 2017, \mnras, 469, 930, \dodoi{10.1093/mnras/stx910}

\bibitem[{Collaboration {et~al.}(2024)Collaboration, the Virgo~Collaboration, \& the KAGRA~Collaboration}]{collaboration2024observation}
Collaboration, T. L.~S., the Virgo~Collaboration, \& the KAGRA~Collaboration. 2024, Observation of Gravitational Waves from the Coalescence of a $2.5-4.5~M_\odot$ Compact Object and a Neutron Star.
\newblock \doarXiv{2404.04248}

\bibitem[{{D'Orazio} \& {Di Stefano}(2020)}]{daniel20}
{D'Orazio}, D.~J., \& {Di Stefano}, R. 2020, \mnras, 491, 1506, \dodoi{10.1093/mnras/stz3086}

\bibitem[{{Dyson} {et~al.}(1920){Dyson}, {Eddington}, \& {Davidson}}]{1920RSPTA.220..291D}
{Dyson}, F.~W., {Eddington}, A.~S., \& {Davidson}, C. 1920, Philosophical Transactions of the Royal Society of London Series A, 220, 291, \dodoi{10.1098/rsta.1920.0009}

\bibitem[{{Ferdman} {et~al.}(2020){Ferdman}, {Freire}, {Perera}, {Pol}, {Camilo}, {Chatterjee}, {Cordes}, {Crawford}, {Hessels}, {Kaspi}, {McLaughlin}, {Parent}, {Stairs}, \& {van Leeuwen}}]{ferdman20}
{Ferdman}, R.~D., {Freire}, P.~C.~C., {Perera}, B.~B.~P., {et~al.} 2020, Nature, 583, 211, \dodoi{10.1038/s41586-020-2439-x}

\bibitem[{Fragione {et~al.}(2018)Fragione, Ginsburg, \& Kocsis}]{Fragione_2018}
Fragione, G., Ginsburg, I., \& Kocsis, B. 2018, The Astrophysical Journal, 856, 92, \dodoi{10.3847/1538-4357/aab368}

\bibitem[{{Goodwin} {et~al.}(1998){Goodwin}, {Gribbin}, \& {Hendry}}]{goodwin98}
{Goodwin}, S.~P., {Gribbin}, J., \& {Hendry}, M.~A. 1998, The Observatory, 118, 201

\bibitem[{{Gould}(1995)}]{gould95}
{Gould}, A. 1995, \apj, 446, 541, \dodoi{10.1086/175812}

\bibitem[{{Han} {et~al.}(2021){Han}, {Wang}, {Wang}, {Wang}, {Zhou}, {Sun}, {Yan}, {Su}, {Jing}, {Chen}, {Gao}, {Hou}, {Xu}, {Lee}, {Wang}, {Jiang}, {Xu}, {Yan}, {Gan}, {Guan}, {Huang}, {Jiang}, {Li}, {Men}, {Sun}, {Wang}, {Wang}, {Wang}, {Xie}, {Xu}, {Yao}, {You}, {Yu}, {Yuan}, {Yuen}, {Zhang}, \& {Zhu}}]{han21}
{Han}, J.~L., {Wang}, C., {Wang}, P.~F., {et~al.} 2021, Research in Astronomy and Astrophysics, 21, 107, \dodoi{10.1088/1674-4527/21/5/107}

\bibitem[{{Hewish} {et~al.}(1968){Hewish}, {Bell}, {Pilkington}, {Scott}, \& {Collins}}]{hewish68}
{Hewish}, A., {Bell}, S.~J., {Pilkington}, J.~D.~H., {Scott}, P.~F., \& {Collins}, R.~A. 1968, Nature, 217, 709, \dodoi{10.1038/217709a0}

\bibitem[{{Hu} {et~al.}(2020){Hu}, {D'Orazio}, {Haiman}, {Smith}, {Snios}, {Charisi}, \& {Di Stefano}}]{2020MNRAS.495.4061H}
{Hu}, B.~X., {D'Orazio}, D.~J., {Haiman}, Z., {et~al.} 2020, \mnras, 495, 4061, \dodoi{10.1093/mnras/staa1312}

\bibitem[{{Hurley-Walker} {et~al.}(2022){Hurley-Walker}, {Zhang}, {Bahramian}, {McSweeney}, {O'Doherty}, {Hancock}, {Morgan}, {Anderson}, {Heald}, \& {Galvin}}]{hurley22}
{Hurley-Walker}, N., {Zhang}, X., {Bahramian}, A., {et~al.} 2022, Nature, 601, 526, \dodoi{10.1038/s41586-021-04272-x}

\bibitem[{{Hurley-Walker} {et~al.}(2023){Hurley-Walker}, {Rea}, {McSweeney}, {Meyers}, {Lenc}, {Heywood}, {Hyman}, {Men}, {Clarke}, {Coti Zelati}, {Price}, {Horv{\'a}th}, {Galvin}, {Anderson}, {Bahramian}, {Barr}, {Bhat}, {Caleb}, {Dall'Ora}, {de Martino}, {Giacintucci}, {Morgan}, {Rajwade}, {Stappers}, \& {Williams}}]{hurley23}
{Hurley-Walker}, N., {Rea}, N., {McSweeney}, S.~J., {et~al.} 2023, Nature, 619, 487, \dodoi{10.1038/s41586-023-06202-5}

\bibitem[{{Ingram} {et~al.}(2021){Ingram}, {Motta}, {Aigrain}, \& {Karastergiou}}]{2021MNRAS.503.1703I}
{Ingram}, A., {Motta}, S.~E., {Aigrain}, S., \& {Karastergiou}, A. 2021, \mnras, 503, 1703, \dodoi{10.1093/mnras/stab609}

\bibitem[{{Katz}(2022)}]{katz22}
{Katz}, J.~I. 2022, \apss, 367, 108, \dodoi{10.1007/s10509-022-04146-2}

\bibitem[{{Kawahara} {et~al.}(2018){Kawahara}, {Masuda}, {MacLeod}, {Latham}, {Bieryla}, \& {Benomar}}]{kawahara18}
{Kawahara}, H., {Masuda}, K., {MacLeod}, M., {et~al.} 2018, \aj, 155, 144, \dodoi{10.3847/1538-3881/aaaaaf}

\bibitem[{{Konstantinidis} {et~al.}(2013){Konstantinidis}, {Amaro-Seoane}, \& {Kokkotas}}]{kons13}
{Konstantinidis}, S., {Amaro-Seoane}, P., \& {Kokkotas}, K.~D. 2013, \aap, 557, A135, \dodoi{10.1051/0004-6361/201219620}

\bibitem[{Kramer {et~al.}(2006)Kramer, Stairs, Manchester, McLaughlin, Lyne, Ferdman, Burgay, Lorimer, Possenti, D'Amico, Sarkissian, Hobbs, Reynolds, Freire, \& Camilo}]{kramer06}
Kramer, M., Stairs, I.~H., Manchester, R.~N., {et~al.} 2006, Science, 314, 97, \dodoi{10.1126/science.1132305}

\bibitem[{{Kruse} \& {Agol}(2014)}]{kruse14}
{Kruse}, E., \& {Agol}, E. 2014, Science, 344, 275, \dodoi{10.1126/science.1251999}

\bibitem[{{Kuang} {et~al.}(2022){Kuang}, {Zang}, {Jung}, {Udalski}, {Yang}, {Mao}, {Albrow}, {Chung}, {Gould}, {Han}, {Hwang}, {Ryu}, {Shin}, {Shvartzvald}, {Yee}, {Cha}, {Kim}, {Kim}, {Kim}, {Lee}, {Lee}, {Lee}, {Park}, {Pogge}, {Mr{\'o}z}, {Skowron}, {Poleski}, {Szyma{\'n}ski}, {Soszy{\'n}ski}, {Pietrukowicz}, {Koz{\l}owski}, {Ulaczyk}, {Rybicki}, {Iwanek}, {Wrona}, {Gromadzki}, {Wang}, {Huang}, \& {Zhu}}]{kuang22}
{Kuang}, R., {Zang}, W., {Jung}, Y.~K., {et~al.} 2022, \mnras, 516, 1704, \dodoi{10.1093/mnras/stac2315}

\bibitem[{{Liu} {et~al.}(2014){Liu}, {Eatough}, {Wex}, \& {Kramer}}]{liu14}
{Liu}, K., {Eatough}, R.~P., {Wex}, N., \& {Kramer}, M. 2014, \mnras, 445, 3115, \dodoi{10.1093/mnras/stu1913}

\bibitem[{Loeb \& Maoz(2022)}]{loeb22}
Loeb, A., \& Maoz, D. 2022, Research Notes of the AAS, 6, 27, \dodoi{10.3847/2515-5172/ac52f1}

\bibitem[{{Lomiashvili} \& {Lyutikov}(2014)}]{lomia14}
{Lomiashvili}, D., \& {Lyutikov}, M. 2014, \mnras, 441, 690, \dodoi{10.1093/mnras/stu564}

\bibitem[{Lorimer \& Kramer(2005)}]{lorimer12}
Lorimer, D.~R., \& Kramer, M. 2005, Handbook of pulsar astronomy, Vol.~4 (Cambridge university press)

\bibitem[{MacLeod {et~al.}(2016)MacLeod, Trenti, \& Ramirez-Ruiz}]{MacLeod_2016}
MacLeod, M., Trenti, M., \& Ramirez-Ruiz, E. 2016, The Astrophysical Journal, 819, 70, \dodoi{10.3847/0004-637x/819/1/70}

\bibitem[{{Maeder}(1973)}]{meader73}
{Maeder}, A. 1973, \aap, 26, 215

\bibitem[{Mandel {et~al.}(2008)Mandel, Brown, Gair, \& Miller}]{mandel08}
Mandel, I., Brown, D.~A., Gair, J.~R., \& Miller, M.~C. 2008, The Astrophysical Journal, 681, 1431–1447, \dodoi{10.1086/588246}

\bibitem[{{Mao} \& {Paczynski}(1991)}]{mao91}
{Mao}, S., \& {Paczynski}, B. 1991, \apjl, 374, L37, \dodoi{10.1086/186066}

\bibitem[{{Marsh}(2001)}]{marsh01}
{Marsh}, T.~R. 2001, \mnras, 324, 547, \dodoi{10.1046/j.1365-8711.2001.04293.x}

\bibitem[{{Masuda} {et~al.}(2019){Masuda}, {Kawahara}, {Latham}, {Bieryla}, {Kunitomo}, {MacLeod}, \& {Aoki}}]{masuda19}
{Masuda}, K., {Kawahara}, H., {Latham}, D.~W., {et~al.} 2019, \apjl, 881, L3, \dodoi{10.3847/2041-8213/ab321b}

\bibitem[{{Montero-Dorta} \& {Prada}(2009)}]{mont09}
{Montero-Dorta}, A.~D., \& {Prada}, F. 2009, \mnras, 399, 1106, \dodoi{10.1111/j.1365-2966.2009.15197.x}

\bibitem[{{Nampalliwar} {et~al.}(2013){Nampalliwar}, {Price}, {Creighton}, \& {Jenet}}]{nampa13}
{Nampalliwar}, S., {Price}, R.~H., {Creighton}, T., \& {Jenet}, F.~A. 2013, \apj, 778, 145, \dodoi{10.1088/0004-637X/778/2/145}

\bibitem[{{Oscoz} {et~al.}(1997){Oscoz}, {Goicoechea}, {Mediavilla}, \& {Buitrago}}]{oscoz97}
{Oscoz}, A., {Goicoechea}, L.~J., {Mediavilla}, E., \& {Buitrago}, J. 1997, \mnras, 285, 413, \dodoi{10.1093/mnras/285.2.413}

\bibitem[{{Paczynski}(1986)}]{pacz86}
{Paczynski}, B. 1986, \apj, 304, 1, \dodoi{10.1086/164140}

\bibitem[{{Paczynski}(1996)}]{pacz96}
---. 1996, \araa, 34, 419, \dodoi{10.1146/annurev.astro.34.1.419}

\bibitem[{{Peters} \& {Mathews}(1963)}]{peters63}
{Peters}, P.~C., \& {Mathews}, J. 1963, Phys. Rev., 131, 435, \dodoi{10.1103/PhysRev.131.435}

\bibitem[{{Pol} {et~al.}(2019){Pol}, {McLaughlin}, \& {Lorimer}}]{pol19}
{Pol}, N., {McLaughlin}, M., \& {Lorimer}, D.~R. 2019, \apj, 870, 71, \dodoi{10.3847/1538-4357/aaf006}

\bibitem[{{P{\"o}ssel}(2019)}]{possel19}
{P{\"o}ssel}, M. 2019, arXiv e-prints, arXiv:2001.00229, \dodoi{10.48550/arXiv.2001.00229}

\bibitem[{{Rahvar} {et~al.}(2011){Rahvar}, {Mehrabi}, \& {Dominik}}]{rahvar11}
{Rahvar}, S., {Mehrabi}, A., \& {Dominik}, M. 2011, \mnras, 410, 912, \dodoi{10.1111/j.1365-2966.2010.17490.x}

\bibitem[{{Ransom} {et~al.}(2014){Ransom}, {Stairs}, {Archibald}, {Hessels}, {Kaplan}, {van Kerkwijk}, {Boyles}, {Deller}, {Chatterjee}, {Schechtman-Rook}, {Berndsen}, {Lynch}, {Lorimer}, {Karako-Argaman}, {Kaspi}, {Kondratiev}, {McLaughlin}, {van Leeuwen}, {Rosen}, {Roberts}, \& {Stovall}}]{ransom14}
{Ransom}, S.~M., {Stairs}, I.~H., {Archibald}, A.~M., {et~al.} 2014, Nature, 505, 520, \dodoi{10.1038/nature12917}

\bibitem[{{Rea} {et~al.}(2023){Rea}, {Hurley-Walker}, {Pardo-Araujo}, {Ronchi}, {Graber}, {Coti Zelati}, {De Martino}, {Bahramian}, {McSweeney}, {Galvin}, {Hyman}, \& {Dall'Ora}}]{rea23}
{Rea}, N., {Hurley-Walker}, N., {Pardo-Araujo}, C., {et~al.} 2023, arXiv e-prints, arXiv:2307.10351, \dodoi{10.48550/arXiv.2307.10351}

\bibitem[{Schneider {et~al.}(1999)Schneider, Ehlers, \& Falco}]{schneider92}
Schneider, P., Ehlers, J., \& Falco, E. 1999, Gravitational Lenses, Astronomy and Astrophysics Library (Springer).
\newblock \url{https://books.google.com.hk/books?id=sPAIgy9QGBsC}

\bibitem[{Shapiro(1964)}]{shapiro64}
Shapiro, I.~I. 1964, \prl, 13, 789, \dodoi{10.1103/PhysRevLett.13.789}

\bibitem[{{Stairs}(2004)}]{stairs04}
{Stairs}, I.~H. 2004, Science, 304, 547, \dodoi{10.1126/science.1096986}

\bibitem[{Suvorov \& Melatos(2023)}]{suv23}
Suvorov, A.~G., \& Melatos, A. 2023, \mnras, 520, 1590, \dodoi{10.1093/mnras/stad274}

\bibitem[{{Tong}(2023)}]{tong23}
{Tong}, H. 2023, \apj, 943, 3, \dodoi{10.3847/1538-4357/aca7fa}

\bibitem[{{Trimble} \& {Thorne}(1969)}]{1969ApJ...156.1013T}
{Trimble}, V.~L., \& {Thorne}, K.~S. 1969, \apj, 156, 1013, \dodoi{10.1086/150032}

\bibitem[{{Wang} {et~al.}(2009{\natexlab{a}}){Wang}, {Creighton}, {Price}, \& {Jenet}}]{wang09b}
{Wang}, Y., {Creighton}, T., {Price}, R.~H., \& {Jenet}, F.~A. 2009{\natexlab{a}}, \apj, 705, 1252, \dodoi{10.1088/0004-637X/705/2/1252}

\bibitem[{{Wang} {et~al.}(2009{\natexlab{b}}){Wang}, {Jenet}, {Creighton}, \& {Price}}]{wang09}
{Wang}, Y., {Jenet}, F.~A., {Creighton}, T., \& {Price}, R.~H. 2009{\natexlab{b}}, \apj, 697, 237, \dodoi{10.1088/0004-637X/697/1/237}

\bibitem[{{Witt} \& {Mao}(1994)}]{witt94}
{Witt}, H.~J., \& {Mao}, S. 1994, \apj, 430, 505, \dodoi{10.1086/174426}

\end{thebibliography}
\bibliographystyle{aasjournal}

\end{CJK*}
\end{document}